\documentclass{article}
\DeclareUnicodeCharacter{2212}{\ensuremath{-}}
\DeclareUnicodeCharacter{2013}{--}
\DeclareUnicodeCharacter{2009}{\,}
\usepackage{preprint}
\usepackage{upgreek}				
\usepackage[utf8]{inputenc} 		
\usepackage[T1]{fontenc}    		
\usepackage[dvipsnames]{xcolor}	
\usepackage{hyperref}       		
\hypersetup{
    colorlinks = true,
    linkcolor = {NavyBlue},		
    anchorcolor = .,
    citecolor = {NavyBlue},		
    filecolor = .,				
    menucolor = .,
    runcolor = {cyan}, 
    urlcolor = {NavyBlue},		
    }
\usepackage{url}            
\usepackage{booktabs}       
\usepackage{amsfonts}       
\usepackage{nicefrac}       
\usepackage{microtype}      
\usepackage{fancyhdr}       
\usepackage{graphicx}       
\usepackage{setspace} 
\usepackage{soul} 
\usepackage{ragged2e} 
\usepackage[toc,page]{appendix}
\usepackage{longtable}
\usepackage{eso-pic}
\usepackage{tikz}
\usepackage[superscript]{cite}

\graphicspath{{./Figures/}}     
\newcommand{\beginsupplement}{%
        \setcounter{table}{0}
        \renewcommand{\thetable}{S\arabic{table}}%
        \setcounter{figure}{0}
        \renewcommand{\thefigure}{S\arabic{figure}}%
     }

  \centering
\title{Sprouting Fields of Conducting Polymer Dendrites as Evolving Organic Electrochemical Frameworks to Interconnect Electrode Arrays}

\author{
  Antoine Baron\textsuperscript{a}, 
  Ewelina Cechosz\textsuperscript{b},
  Konrad Szaci{\l}owski\textsuperscript{b},
  Enrique H. Balaguera\textsuperscript{c},\\
  \textbf{S\'{e}bastien Pecqueur\textsuperscript{a}}  \\
  \\
  a. IEMN, UMR 8520 \\
  Univ. Lille, CNRS, Univ. Polytechnique Hauts-de-France\\
  59000 Lille, France\\
  \\
  b. Academic Centre for Materials and Nanotechnology\\
  AGH University of Krakow\\
  Krak\'{o}w, Poland\\
  \\
  c. Escuela Superior de Ciencias Experimentales y Tecnolog\'{i}a, \\ Universidad Rey Juan Carlos, C/ Tulip\'{a}n, \\
  28933 M\'{o}stoles, Madrid, Spain\\
  \\
  \texttt{sebastien.pecqueur@iemn.fr} \\
}

\begin{document}
\maketitle

\begin{abstract}
Balancing high performance with sustainability in microelectronics is challenged by energy-intensive processes, scarce materials and limited recyclability.
Inspired by nature, a promising solution relies on the evolution of the foundry model to transient microelectronics, using electrochemistry to grow adaptive circuits on demand and during operation.
To this end, this study explores how to harness the morphogenesis of conducting polymer dendrites between stacks of micro-electrode arrays to dynamically interconnect them with voltage pulses.
Using propylene carbonate as a solvent, we investigated the dynamics of the electroconvective growth of poly(3,4-ethylenedioxythiophene) (PEDOT), revealing strong structural dependencies between the morphology of the electrogenerated polymer and the confinement of the micro-electrodes.
As the system evolves from 2D flat patterns to 3D dendritic structures, the capability of the conducting polymer to locally interconnect specific nodes heavily depends on the size and density of the arrays which directly impact the flow distribution during growth.
By unraveling how a device geometry enables the growth of dendritic interconnections on integrated micro-electrode arrays, the morphogenesis of conducting polymer dendrites not only addresses the longstanding issue of heterogeneous integration in microelectronics, but also opens the way to a paradigm where electronic devices learn as they grow in our pockets.

\end{abstract}

\raggedright
\keywords{Parallel interconnectivity \and Evolving electronics \and Sustainable microelectronics}

\newpage 

\justifying

\section{Introduction}
Balancing high performance with sustainable manufacturing is a challenge.\cite{Hermundsdottir2021,Hermundsdottir2022,AhmadiGh2022} 
In microelectronics, this need is particularly acute, as foundries rely on high energy consumption, vast supplies of water, the availability of rare elements and the recycling of chemical waste.\cite{Villard2015,Kuo2022,Ruberti2023,IEEE_ESHS_2024,Sandhu2025} 
With the ever-growing demand for chips, driven by the emergence of artificial intelligence in many sectors, a global shift toward a new paradigm is urgently required.\cite{IEEE_ESHS_2024,Lanza2025}\\[3pt]
Biological intelligence is a lasting source of inspiration finding new ways to build systems that process information with limited resources via low-energetic processes.\cite{Tero2010,Kar2016,Hassabis2017,Moro2022,Mehonic2022,Boufidis2025}\\[3pt]  
A global observation is that, unlike nature, we traditionally separate the constraints for manufacturing and operation: The energy allocated to build electronics is not used for processing, and vice versa.
An important consequence is that energy-efficient technologies are usually benchmarked on the energy cost of a single elementary task operation,\cite{Ionescu2011,Burgt2017,Zhai2021,Demasius2021} while overlooking the energy allocated to fabricate and recycle the hardware:\cite{Sustain2026} energy that when accounted for in a life cycle assessment,\cite{Villard2015,Kuo2022,Ruberti2023,Liu2024} contributes significantly to the overall environmental cost. Furthermore, the complexity of the supply chain of rare coinage elements is usually neglected.\cite{Ekatpure2022}\\[3pt]
Typically, conventional electronic systems are assembled once and maintained passively, whereas biological systems integrate evolution as part of both their development and their adaptation to an environment. In nature, the growing process itself may act as a medium for information processing: concentration changes, morphology, geometry, or physicochemical parameters serve simultaneously as inputs and outputs\cite{De2018,Heisenberg2013,deSilva2007}. When exposed to external stimuli, such as pH changes or light, chemical systems can alter their growth kinetics or bond formation~\cite{De2018}. Consequently, the chemical signal shapes the architecture of the system, while the architecture, in turn, determines its properties. An example of such a feedback loop is autocatalysis, in which a reaction product accelerates its own formation, enabling signal amplification and kinetic memory~\cite{Baltussen2024}. In open systems, a continuous exchange of energy or matter can sustain a non-equilibrium state, in which the reaction product is not a static state but a dynamically evolving structure\cite{De2018,Needleman2017}.
The relationship between the generation of form and its function is central to morphogenesis: complex architectures emerge through local interactions, growth and symmetry breaking rather than being fully specified in advance.\cite{pismen2020}

Inspired by the behavior of animal brains, plant roots or fungal mycelium, evolving electronics propose to relax manufacturing constraints by allowing electric systems to change their structure during operation.\cite{Yao1999,Zebulum2001,Greenwood2006,Greenwood2010,Haddow2011,Greenwood2015}
In such systems, specific connections in the hardware form dynamically to optimize its functionality: For example, a system could increase specific resources in its processing unit if complex computational tasks require it, or heal from physical damages incurred by such tasks in a real-world environment. Conducting polymer filaments offer a promising material platform for this approach as their nucleation, branching, growth and interconnection can generate complex conductive networks whose topology determines the distribution of electrical signals. When electrical stimulation modifies filament growth or the internal state of the network, the coupling between electrical activity and material evolution could provide a route to adaptive, history-dependent computation.
By addressing decentralized manufacturing and retail shortages but also obsolescence and recyclability, letting electronics to grow autonomously could  unlock unprecedented information processing potentials to feature micro-electronic devices in a system.\\[3pt]
To enable this, the covalent bonds of silicon atoms on a wafer are too hard to break, while essential to ensure stability for conventional electronics. 
For applications requiring material evolution, operational substrates must facilitate mass transfer at a low energetic cost, powered by electricity. 
Solid state electrochemistry can enable material diffusion at the atomic level, but understanding how to transfer from hardware electronics to wetware iontronics is the true challenge to tackle for unlocking evolutionary technologies.\\[3pt]
Continuous progress is being made in dynamic electrochemical systems in liquids, as their functional architecture can emerge, reconfigure, or regenerate during operation rather than being fully defined during device fabrication.\cite{Robin2023,Ismail2025}
Evolving liquid systems include nanofluidic ionic memristors, in which the conductance of an aqueous electrolyte confined in a nanochannel depends on the history of applied potentials, ion concentrations, and ion transport, owing to interfacial processes such as ion adsorption/desorption and surface-charge regulation.\cite{Robin2023,Ismail2025,Kamsma2024} 
Commonly employed architectures include two-dimensional channels based on materials such as hexagonal boron nitride (hBN), filled with aqueous salt electrolytes.\cite{Robin2023,Ismail2025}
Because ions serve as charge carriers, as in biological neurons, these devices couple ionic signals directly to electronic readout at the electrodes; accordingly, they exhibit synapse-like plasticity and find applications in neuromorphic and reservoir computing.\cite{Robin2023,Kamsma2024}\\[3pt]
Electronic circuits based on liquid-metal droplets, typically eutectic gallium--indium have also been proposed for self-healing and flexible electronics.\cite{Dickey2017,Markvicka2018,Sunwoo2025} 
In such systems, conductive pathways can be formed under external stimuli and mechanical deformation through changes in the arrangement of the droplets.\cite{Markvicka2018,Sunwoo2025}
The fluidic nature of the metallic phase allows continuous rearrangement of conductive pathways without irreversible fracture, while its low electrical resistance and high mechanical resilience promote robust operation. 
Nevertheless, long-term stability remains challenging owing to surface oxidation, associated with the formation of a native Ga\textsubscript{2}O\textsubscript{3} skin, and wettability effects.\cite{Dickey2017,Sunwoo2025}
One strategy to address this limitation is the development of composite materials in which liquid-metal droplets are embedded within elastic matrices such as silicone elastomers, for example polydimethylsiloxane (PDMS);\cite{Markvicka2018} such matrices mechanically stabilize the liquid phase while preserving its ability to deform, reconnect, and restore electrical continuity after damage.\cite{Markvicka2018}\\[3pt]
Among the operating principles of related liquid-phase information-processing systems are redox reaction networks driven by chemical inputs or light exposure, such as photosensitive Belousov--Zhabotinsky-type oscillators.\cite{Kdr1997}
In chemical reservoir computers, information is encoded in time-dependent concentration profiles, oxidation states, and reaction pathways rather than in a fixed circuit topology.\cite{Kheirabadi2023,Baltussen2024}
Examples include self-organizing reaction networks such as the formose reaction, which has been shown to perform nonlinear classification, dynamical-system emulation, and time-series forecasting.\cite{Baltussen2024}
Although such systems offer a promising route toward parallel and adaptive chemical information processing, their slow reaction dynamics and limited process controllability remain significant constraints. 
Further progress therefore requires improved control over reaction kinetics, mass transport, and the reproducibility of emergent conductive or optical structures.\cite{Kheirabadi2023,Robin2023,Kamsma2024,Baltussen2024}\\[3pt]
In this perspective, conducting polymer dendrites (CPDs) have demonstrated a great potential to enable evolving electronics.\cite{Baron2024,Baron2024a} 
In organic and aqueous solutions, conducting polymers grow electrodynamically, forming connections over millimeter distances between electrodes.\cite{Koizumi2016,Ohira2017,Watanabe2018}
As both an ion and an electron conductor,\cite{Rivnay2016,Paulsen2020} the poly(3,4-ethylenedioxythiophene) (PEDOT) polymer, stable in its p-doped state,\cite{Chiu2006,Crispin2006} behaves as an electrochemical transistor rather than a metal:\cite{Inal2017,Tropp2023} the material is electrically programmable. 
The topological richness of these objects introduces dynamical complexity that can be exploited for computation, particularly in recurrent neural networks.\cite{Cucchi2021,Petrauskas2021,Scholaert2025} 
Different topologies can be induced either electrically or chemically.\cite{Janzakova2021,Baron2026}
Consequently, the development of such systems can be triggered solely by an electrical signal carrying information to encode,\cite{Baron2025} fueled by chemical energy stored in the fluid as electroactive precursors. 
Results demonstrating hydrodynamical control via the fluid viscosity confirm that a continuous liquid phase is required to promote charged particle motion necessary for CPD morphogenesis.\cite{Janzakova2021a,Kumar2022,Baron2026,Baron2026a}\\[3pt]
One of the main limitations restricting the  applicability of CPDs for device integration is managing solubility and volatility. 
While water is a moderately volatile solvent, the limited solubility of the EDOT monomer in this medium works against the growth of CPDs.\cite{Baron2026}
In organic solvents like acetonitrile, EDOT is more soluble, allowing more precursors to fuel growth between planar electrodes,\cite{AkaiKasaya2020,Cucchi2021,Petrauskas2021,Hagiwara2021,Watanabe2024} but its volatility limits system operability to a very small time window.\\[3pt]
To overcome these limitations, this study uses propylene carbonate as a solvent. 
As a non-volatile solvent widely employed in lithium-ion batteries and skincare,\cite{Fayaz2024,Rahman2025} it is an ideal candidate for implementing electropolymerization on a chip.\cite{Moustiez2025} 
Here, its solubility enables the growth of three-dimensional (3D) CPD networks, with both vertically aligned and in-plane configurations, effectively connecting distinct layers of physical electrical contacts. 
This approach not only addresses the challenge of parallel interconnectivity but also paves the way for bio-inspired, eco-friendly electronic architectures.\\[3pt]

\newpage

\section{Experimental}

\subsection{Materials and Methods}
All the chemicals used in this study were purchased from commercial vendors and utilized without further treatment. 
All manipulations were performed under ambient conditions (room temperature and atmospheric air).
3,4-Ethylenedioxythiophene (EDOT), propylene carbonate (PC), poly(ethylene glycol)-block-poly(propylene glycol)-block-poly(ethylene glycol) (P123) and p-benzoquinone (BQ) were obtained from Sigma-Aldrich. 
1-Ethyl-3-methylimidazolium trifluoromethanesulfonate (EmimOTf) was purchased from Solvionic. 

\subsection{Device Microfabrication}
This study involves at least three experimental setups. 
A first consists of two gold wires (25~$\upmu$m in diameter) immersed in a solution droplet; it is a setup identical to the one described in previous studies,\cite{Janzakova2021} except for the composition of the droplet which involves propylene carbonate as the solvent here. 
A second setup uses two coplanar 25-$\upmu$m wide gold electrodes, lithographed on the same SiO\textsubscript{2} substrate. 
A third setup involves two planar electrodes, each consisting of a 1~mm-thick 4.9~cm~$\times$~4.9~cm glass substrate. 
The bottom electrode was coated with a 100~nm gold layer deposited on a 10~nm titanium adhesion layer by e-beam evaporation using a PLASSYS MEB 550SL, while the top electrode was coated with only 12~nm of gold for 3~nm of titanium, to render it semitransparent and allow optical observation of the growth from above. 
The bottom electrode is coated with a 2~$\upmu$m thick Parylene C passivation layer, deposited by chemical vapor deposition. 
Hole openings in the Parylene C were etched using O\textsubscript{2} plasma reactive-ion etching. 
Both planar electrodes were similarly coated with Parylene C and structured following the same protocol.

\subsection{Electropolymerization and Characterization}
Different electropolymerization signals were used to create diverse CPD morphologies on electrodes, with parameters further detailed for each case. 
Unless specified otherwise, the electropolymerization signal consisted in a unipolar alternating signal to oxidize EDOT only on the bottom planar electrode. 
In all cases, the signal was generated using a Modulab XM MTS (Solartron Analytical) impedance analyzer. 
The dynamical growth signals are applied between the two electrodes at a given frequency \textit{f} and duty cycle dc. 

\subsection{Image Capture and Processing}
The growth is observed from above the top electrode using a VGA CCD color Camera from HITACHI Kokusai Electric Inc. 
Following an ethanol rinse and drying under a gentle stream of N\textsubscript{2}, the patterns and CPDs presented in this study were all captured using a Keyence VHX-6000 microscope or an Ultra 55 SEM from Zeiss. 
Image processing to determine characteristic sizes was performed using an ImageJ script, involving a combination of contrast enhancement, thresholding and watershed separation.
  
\newpage

\section{Results}

\subsection{Conducting Polymer Dendrite Morphogenesis in Propylene Carbonate}

\begin{figure}[!h]
  \centering
  \includegraphics[width=1\columnwidth]{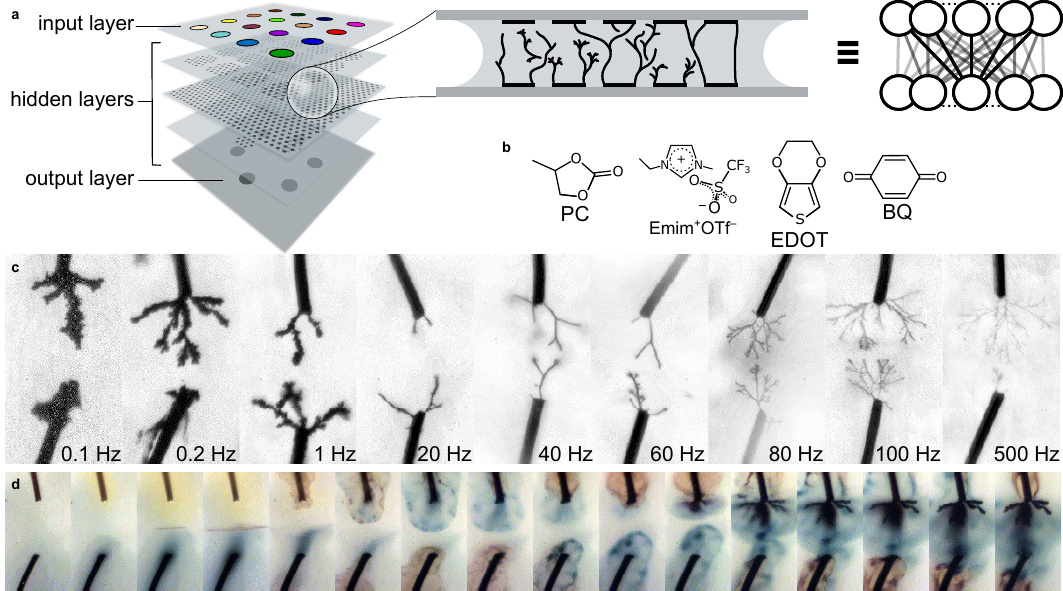}
  \caption{\textbf{Conducting Polymer Dendrites Single Growth in Propylene Carbonate $\vert$}
  \textbf{a,} The target device architecture, justifying the use of a non-volatile organic solvent for the electroactive solution.
  To implement CPD morphogenesis to electrically interconnect planes of 2D electrode array, precursors must be sufficiently concentrated in a confined liquid environment, with a steady content during the experiments.
  \textbf{b,} Chemical formulae of the molecules contained in the solution. 
  \textbf{c,} Microscope pictures taken during several growths showing the dependency of a CPD morphology (100~mM EDOT, 100~mM BQ, 10~mM EmimOTf in PC) with the frequency of an applied voltage square signal (amplitude: 5~V, offset: 0~V), grown with a free-standing gold wire setup. 
  \textbf{d,} Enhanced color contrast of sequential frames showing the growth at \textit{f}~=~0.2~Hz (chronological order from left to right).
  }
  \label{fig:fig1}
  \end{figure}

Previous works focusing on CPDs study their behavior either individually as isolated junctions, or collectively as a network in a 2D plane.\cite{Cucchi2021,Scholaert2025}
The interdependency of CPDs' electrical behavior in a common electrolyte solution relevantly exploits their rich topology for information processing.
Still, confining such topologies on a 2D plane limits the number of interconnections to achieve when integrated into an electrical system.
Although CPDs can exhibit high affinity for a 2D planar substrate,\cite{Watanabe2018} out-of-plane CPD growth has already been achieved by Hagiwara and coworkers at the level of a single junction with lithographically patterned electrodes.\cite{Hagiwara2021}
To assemble many of them in a layered system (Fig.~\ref{fig:fig1}a), the electroactive solution must support the growth of three-dimensional structures.
However, recent results showed that the viscosity of the growth medium is a limiting factor that hinders the drift of conducting polymer particles that occurs during morphogenesis.\cite{Baron2026}
Colloid materials such as gels or porous solids with capillary condensation (such as wet zeolites, metal-organic frameworks) could present advantages to structure them.
However, the complexity of their growth must be understood when limited by the boundaries of a confined cavity.
To do so, this study focuses on the use of propylene carbonate and an ionic liquid, as non-volatile agents (Fig.~\ref{fig:fig1}b).
Provided that CPD growths are highly influenced by the chemical nature of their environment, the electrodynamical behavior of growing CPDs has been studied on free-standing wires at various signal frequencies (Fig.~\ref{fig:fig1}c--d).

Previous experiments have revealed an important effect of the chemical environment on CPD growth\cite{Baron2026}. Specifically, replacing NaPSS with EmimOTf dramatically reduced the growth completion times. In water, with NaPSS, the wires connect each other in around 90 s, while the growth only requires around 15 s to complete with triflate ions. Shorter completion times are also observed in PC with the growths presented in Fig. \ref{fig:fig1}c at different growth frequencies, which suggests that the change of solvent from water to PC may have a minimal effect on the growth dynamics and morphology. The viscosity of PC being approximately three times higher than that of water at 298.15 K\cite{Muhuri2002}, it should have increased the growth duration according to previous results of growths in glycerol/water mixtures\cite{Baron2026}. Rather, it appears that the morphology may be mainly dictated by the nature of the ions.
The same change in branch thickness observed in water with the frequency (larger branches at lower frequencies, thinner branches at higher frequencies\cite{Janzakova2021}) is also observed in PC. Since PC has a wider electrochemical window that increases the range of potentials where gas bubbles will not arise and prevent growth\cite{Sritharan2015}, the growth frequency can be reduced down to frequency values inaccessible in water (which has a lower frequency limit of around 10 Hz for CPD growth\cite{Janzakova2021}). These findings demonstrate that while there is an upper limit to the growth frequency, likely linked to how fast ions can travel and take part in the reactions at the electrodes, there is fundamentally no lower frequency limit for dendrite formation.\\
[3pt]
Electrochromism, the change of color observed upon doping or dedoping a conducting polymer, has never been reported with PEDOT:PSS dendrites grown in water. However, PEDOT is electrochromic as a thin-film in an aqueous electrolyte doped with PSS ions\cite{KAWAHARA2012}. Thus, some feature specific to CPD growth in water, possibly related to their micrometer-scale thickness\cite{Baron2026} or morphology, may prevent a visible color modulation. Similarly, PEDOT:OTf dendrites grown in water also do not display the change of color. Meanwhile, changes in oligomer solubility dependent on the solvent have been shown to affect the morphology of the film and impact the electrochromic properties in PC\cite{Poverenov2010}. When growing PEDOT:OTf CPDs in PC, as shown in Fig. \ref{fig:fig1}d, different colors are observed on both sides: the electrolyte surrounding the anodic wire becomes blue, while the cathodic wire side appears mainly dark red or brown. As the polarity changes for each half period, the colors also change sides every half period. Moreover, during the first half period when no PEDOT is polymerized on the cathodic side, a yellow cloud is produced around the wire. Since PEDOT is known to be strongly absorbing and blue in its neutral state, while its doped state is more transmissive \cite{KAWAHARA2012}, the brown coloration is hypothesized to result from the superposition of reduced dark-blue PEDOT and a yellow contribution. The latter may originate from the redox chemistry of benzoquinone, which is electrochemically active in PC \cite{Ji2006}, as a distinct yellow cloud is already observed at the cathodic wire during the first half-cycle, before any PEDOT has formed on this side. On the anodic side, the light blue color may result from the oxidation of PEDOT. Its electrochromism in PC may therefore be responsible for the color changes through the doping/dedoping of a diffuse cloud of polymer around the wires. After a few cycles, it is hypothesized that local solubility changes may lead to two separate phases, the solid, dendritic one and the cloud of polymer/oligomer that disappears once the voltage is turned off.

\subsection{Growth Boundaries with the Inter-Electrode Confinement and the Substrate Affinity}

\begin{figure}
  \centering
  \includegraphics[width=1\columnwidth]{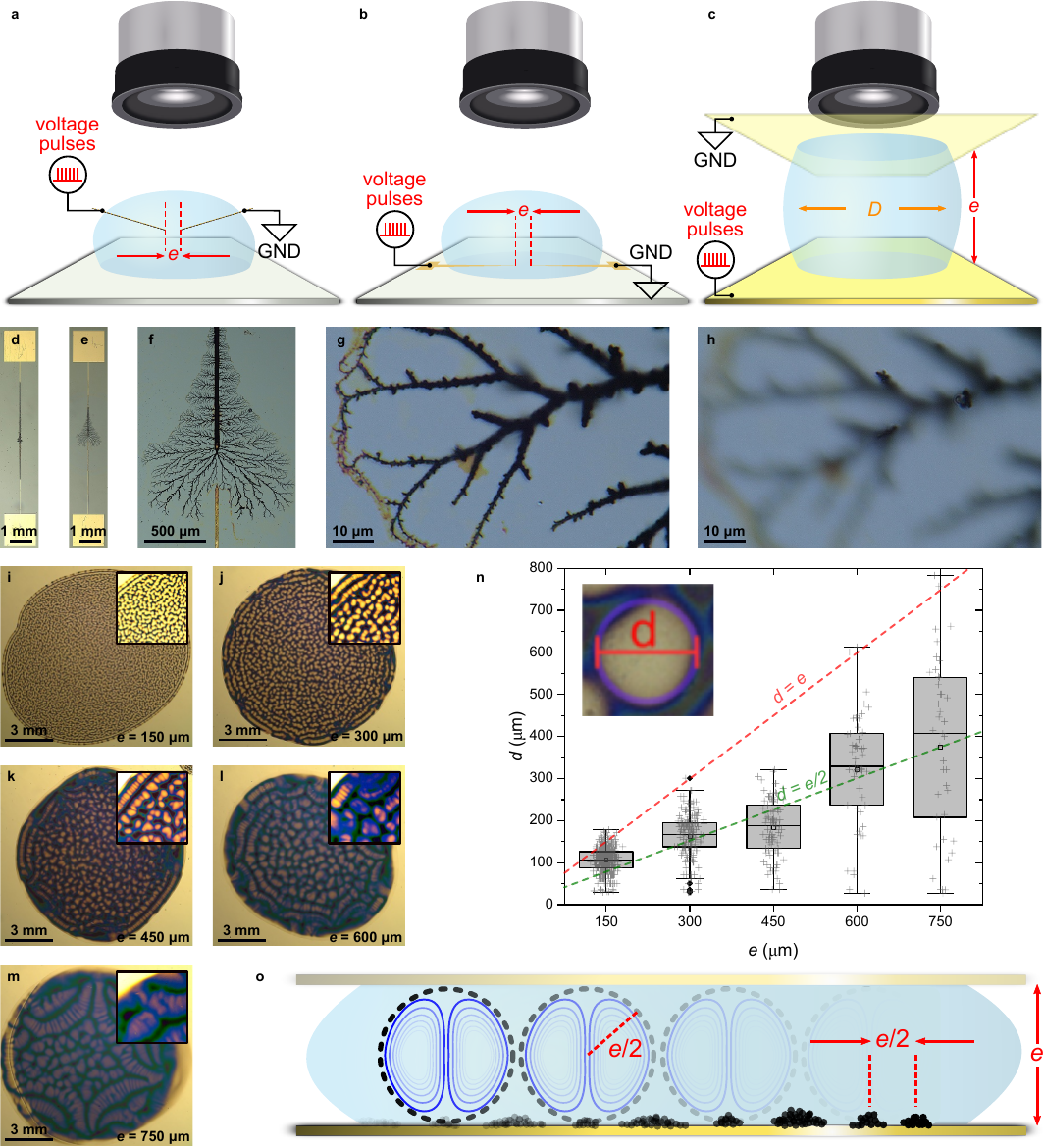}
  \caption{\textbf{The Diversity of PEDOT Structures Observed with Electropolymerization in Different Configurations $\vert$}
  \textbf{a,} Schematic of the free-standing wire setup. 
  \textbf{b,} Schematic of the coplanar electrodes setup. 
  \textbf{c,} Schematic of the parallel plates setup. 
  \textbf{d,} Low-magnification image of the coplanar electrodes with CPDs grown with a symmetric square wave voltage (\textit{f}~= 20~Hz, dc~= 50\%, amplitude~: 5~V). 
  \textbf{e,} Low-magnification image of the coplanar electrodes with CPDs grown under a constant voltage (5~V). 
  \textbf{f,} Close-up view of \textbf{e}. 
  \textbf{g,} High magnification view of \textbf{e} focused on the substrate. 
  \textbf{h,} High magnification view of \textbf{e} focused slightly above the substrate. 
  \textbf{i--m,} Optical microscope pictures of the bottom electrode in the parallel plate setup after electropolymerization  with unipolar voltage pulses (\textit{f}~= 1~kHz, dc~= 50\%, pulse~: 5~V), at different interelectrode distances. 
  \textbf{n,} Plot showing the evolution of the average diameter \textit{d} of the patches against the interelectrode gap. 
  \textbf{o,} Schematic illustrating the electroconvective hypothesis behind the formation of PEDOT patterns in the parallel plate setup.
  }
  \label{fig:fig2}
  \end{figure}

Conducting polymer dendrites are versatile structures that can be grown in various ways using different electrical signals and electrode geometries. 
While the free-standing wire configuration (as depicted in Fig.~\ref{fig:fig2}a) allows studying CPD morphologies in different liquid environments without any interactions with a solid surface,\cite{Janzakova2021,Baron2024,Baron2026} a 2D configuration (as depicted in Fig.~\ref{fig:fig2}b) using lithographically patterned electrodes enables a easier electrical contact at multiple points and allows the system to be exploited as a circuit rather than single device objects. \cite{Cucchi2021,Scholaert2025}
To gain access to the 2.5D configuration (as depicted in Fig.~\ref{fig:fig2}c)\cite{Pecqueur2025}\textsuperscript{,pp.63}, one must first understand the impact of both the substrate as a zero-defect electrode, and the limitation of reactive species in a confined layered environment.\\[3pt]
To assess the affinity of CPDs in PC with a solid substrate, growths have first been studied with the setup pictured in Fig.~\ref{fig:fig2}b with two coplanar electrodes to promote the growth along the solid substrate (Fig.~\ref{fig:fig2}d--h), under various voltage conditions.
As visible in Fig.~\ref{fig:fig2}d, CPDs can grow from coplanar electrodes using PC as the solvent and a symmetric square wave voltage.
With such a voltage signal, both electrodes are intermittently anodes allowing EDOT oxidation over their surface immersed in the PC solution drop. However, the growth eventually extends beyond the substrate/solution interface and bridges the two electrodes.
In the case of a DC voltage polarization, as no bubbling was observed in\ref{fig:fig1} at low frequency, growth can occur but only from one electrode, the anode (Fig.~\ref{fig:fig2}e--f).
Even in such cases, dendrites form Lichtenberg-like figures,\cite{Thomas1945,Takahashi1979,Ficker1999} despite the continuous solid/liquid interface along the growth. 
However, the base of the branches seems to adhere to the surface, suggesting a potential affinity of a CPD for the 2.5D configuration. 
Despite this apparent affinity to the substrate, some secondary branches can develop out of the plane (Fig.~\ref{fig:fig2}f--g), indicating that vertical growth of an oxidizing electrode can coexist with planar polymer growth on the same electrode.\\[3pt] 
When using the two parallel plates as electrodes, the electropolymerized structures are completely different (Fig.~\ref{fig:fig2}h--m).
In such configuration, grown patterns resemble the one obtained with a water-based solution, as recently reported by Baron \textit{et al.}\cite{Baron2026b}
Heterogeneous coatings are deposited on the anode when polarized with unipolar voltage pulses, with no CPDs extending vertically. 
Some differences in the arrangement of the patterns imprinted with PC are noticeable compared to the earlier results performed in an aqueous electrolyte solution.\cite{Baron2026b} 
Specifically, in water, the pattern featured different patches' morphologies depending on the chemical and electrical parameters. 
In pure water and under comparable voltages, the patches were black, surrounded by a bright matrix. 
In Figs. \ref{fig:fig2}i--m, the patches appear inverted, bright and surrounded by a dark matrix. 
When increasing the gap between the two electrodes, the characteristic size of the patches also increases (Fig.~\ref{fig:fig2}l--n): a property which was previously observed with an aqueous electrolyte solution.\cite{Baron2026b} 
An empirical scaling law is found with the size of the electrochemical cavity, as plotted in Fig. \ref{fig:fig2}n: a linear fit reveals that the characteristic patch size \textit{d} remains statistically half the gap width \textit{e}, and rarely exceeds it. 
As previously hypothesized in water,\cite{Baron2026b} the cause of these observed patterns is suspected to be electroconvective vortices agitating the solution during the electropolymerization (illustration in Fig.~\ref{fig:fig2}o), which can lead to higher concentrations in more stable regions (such as between vortices),\cite{Demekhin2014,Valenca2015,Kim2019,Provat2025} whereas the vortices themselves may locally prevent PEDOT deposition.\\[3pt] 
These results confirm the analogy with earlier results obtained in water:\cite{Baron2026b} CPDs can grow in a free standing configuration or along a 2D substrate: the sole presence of a solid/liquid interface is not a strict hindrance to the morphogenesis of CPDs. 
However, severe limitations remain when the solution is confined between two parallel planar electrodes.
The next part shows that these limitations are not strict but highly concentration- and chemistry-dependent.\\

\newpage

\subsection{Higher Morphological Versatility with Wider Voltage and Concentration Conditions}
\begin{figure}[!h]
  \centering
  \includegraphics[width=1\columnwidth]{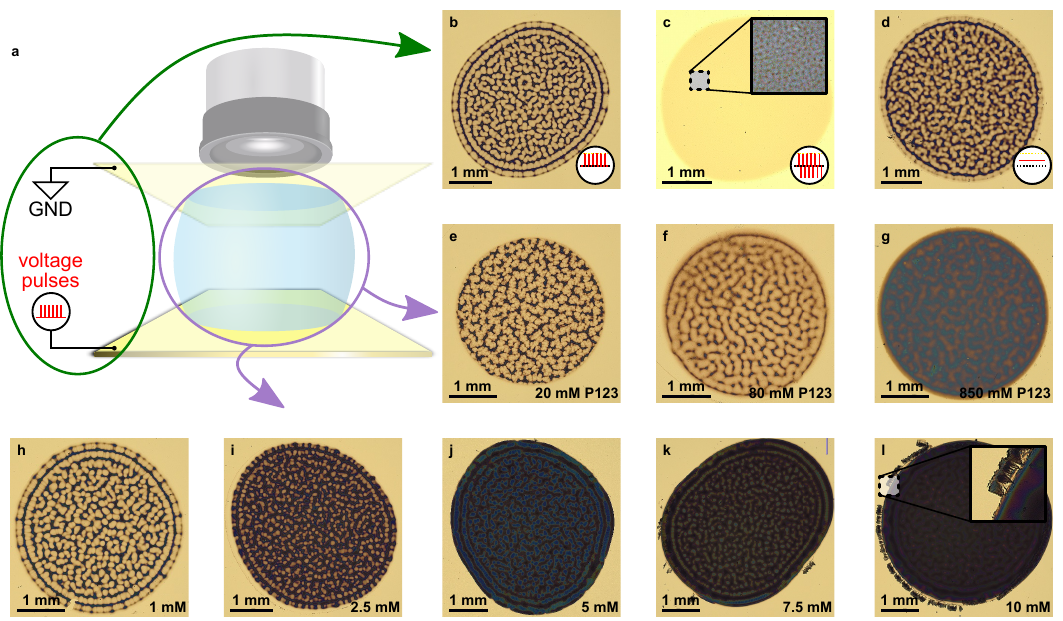}
  \caption{\textbf{Pattern Growth in Propylene Carbonate Depends on Both the Voltage Signal and the Chemical Composition of the Droplet $\vert$} 
  \textbf{a,} Schematic illustrating the two main variables studied in this figure: the nature of the voltage applied to the two electrodes (\textbf{b--d}) and the chemical composition of the droplet (\textbf{e--l}). 
  \textbf{b,} Reference pattern (unipolar voltage, \textit{f}~= 100~Hz, dc~= 50\%, pulse~: 5~V, 1~mM EmimOTf, 100~mM EDOT, 100~mM BQ).
  \textbf{c,} Pattern obtained with the same concentrations, with a bipolar voltage (\textit{f}~= 10~Hz, dc~= 50\%, amplitude~= 5~V, offset~= 0~V). 
  \textbf{d,} Pattern obtained with the same concentrations, with a constant voltage (2.5~V). 
  \textbf{e--g,} Patterns obtained at increasing concentrations of P123 (unipolar voltage, \textit{f}~= 1~kHz, dc~= 50\%, pulse~: 5~V, 1~mM EmimOTf, 100~mM EDOT, 100~mM BQ). 
  \textbf{h--l,} Patterns obtained at increasing concentrations of EmimOTf (unipolar voltage, \textit{f}~= 1~kHz, dc~= 50\%, pulse~: 5~V, 100~mM EDOT, 100~mM BQ).}
  \label{fig:fig3}
  \end{figure}

Working in PC rather than water offers three important advantages: 
The first one is the low volatility of PC,\cite{Haynes2016} allowing to study them for longer time without observing noticeable drifts: a prerequisite to limit experimental artifacts for low frequency impedance characterization or dynamic electrical tests.
The second one is the wider electrochemical window,\cite{Prudlik2025,Chauhan2026} allowing to study the electrochemical growth of PEDOT under high operating voltage and low frequency without formation of bubbles, and complexifying the growth mechanisms with the generation of an uncontrolled fourth phase in addition to the electrode, the polymer and the solution.
The third advantage is the higher solubility of the monomer,\cite{Schweiss2005,Zhang2024} allowing to explore the growth dynamics in confined solutions without being limited by the supply of chemical reagents contained in the solution.
This section explores the impact of the voltage input, the chemicals involved and their concentration, to promote the transition from 2D spot patterns to 3D vertical dendrites and assess if both can coexist to make 2.5D conducting polymer frameworks, as 3D topologies confined in a solution layer (Fig.~\ref{fig:fig3}a).\\[3pt]

While CPDs are usually grown with trains of bipolar pulses between electrodes,\cite{Janzakova2021} the 2D patterns presented in the previous part are obtained by applying unipolar pulses.\cite{Baron2026b} 
This is less a necessity than a choice of convenience, in order to observe growths through the semi-transparent top electrode \textit{in operando} under the microscope without obscuring the top surface with a dark coating of polymer.
To study the coexistence of both polymer formation in common voltage conditions, three different cases are reported in Figs. \ref{fig:fig3}b--d.
2D patterns obtained with unipolar pulses (Fig.~\ref{fig:fig3}b) are very contrasted and disordered.
A substantial difference with the ones obtained in water in a previous study is the double annular organization of spots at the air/solution interface (previous coatings with an aqueous solution always featured one single annular organization).\cite{Baron2026b}
Surprisingly, applying the same bipolar pulses as the one applied on a free-standing wire configuration does not lead either to CPDs nor contrasted spots (Fig.~\ref{fig:fig3}c).
Apparently uniform, we noticed spot patterns after increasing the contrast of the optical microscope images (Fig.~\ref{fig:fig3}c), although the spots in this case have a much shorter wavelength than the ones on the unipolar pattern. 
In fact, it was also observed that this wavelength is dependent on the frequency of an AC signal the same way as the branch thickness of the dendrites: higher frequencies give shorter wavelength for spots, and thin branches for dendrites.
The larger and contrasted spots obtained under unipolar pulses differ substantially from the ones grown under a bipolar voltage. In fact, a growth obtained under unipolar conditions (Fig.~\ref{fig:fig3}b) resembles far less the ones obtained with bipolar pulses (Fig.~\ref{fig:fig3}c) than the ones obtained under a DC voltage (Fig.~\ref{fig:fig3}d).
Highly contrasted spots are obtained without dynamic voltages, showing that the dynamics of an alternating voltage supply is not required to trigger the fluid electroconvection responsible for spot formation. An AC voltage is also not required for growing CPDs, as was observed in one example in Fig.~\ref{fig:fig3}e-f. In fact, a sufficiently low-frequency AC voltage can effectively behave as a DC voltage over the timescale relevant to spot formation (spots can grow within about 10 s, so a 10 s half-period is sufficient to mimic a DC voltage pattern). This suggests that the increase in wavelength with decreasing frequency may saturate below a sufficiently low frequency, corresponding to a maximum vortex size. Such a saturation of the wavelength at low frequency has already been reported for electroconvection in other works\cite{Yossifon_2008_}.
The special case of a unipolar voltage, always growing large and highly contrasted spots despite frequency changes, may be caused by the existence of a DC component in the signal, which could explain why the resulting pattern appears closer to its DC counterpart than its AC counterpart.
The two experimental setups, on one side two anisotropic electrodes, on the other two parallel plates, therefore display a qualitatively consistent dependence on the temporal characteristics of the growth voltage, suggesting similarities between the two growth processes. Spots and CPDs, may, in addition, coexist rather than compete.\\[3pt]
Another parameter that may impact the formation of these patterns is the chemical content of the droplet.
In water, the addition of glycerol changes the organization of spot patterns,\cite{Baron2026b} presumably caused partially by the increased viscosity.\cite{Baron2026}  
In PC, the effect of the viscosity of the solution on the printed pattern has been studied by adding P123 to the solution.
P123 is a block copolymer neutral surfactant,\cite{Wanka2002,PittoBarry2014,Tripathi2025} that is generally used with water to make hydrogels and micellar suspensions.\cite{Chaibundit2007,Chaibundit2008,Vyas2025}
In our experiments, P123 was used in PC as a thickener to increase the solution viscosity.
After adding different concentrations of P123, no gelation was observed over time: we do not suspect the formation of an organogel with PC, but do not rule out the formation of a colloidal environment.
At low P123 content (Fig.~\ref{fig:fig3}e), the imprinted pattern already features differences with a pattern generated in pure PC (Fig.~\ref{fig:fig3}b). 
Notably, the double ring at the periphery of the pattern has disappeared. 
Following the conclusions of a previous study in water, this ring is suspected to be caused by the formation of a continuous convective roll as vortices progressively order themselves parallel to each other at the droplet walls. 
In this case, viscosity appears to prevent this structure. 
No significant change in the wavelength is observed however. 
With an increased concentration of P123, the pattern features a smoother contrast (Fig.~\ref{fig:fig3}f). 
Eventually, at higher P123 concentrations, the formation of horizontal fibers on top of the dark separations between bright spots is observed, similar to a spiderweb (Fig.~\ref{fig:fig3}g). 
This may be explained by the increased viscosity slowing down the vortices in solution, increasing the electropolymerization reaction rate between them.
Despite the fact that no vertical CPDs were obtained in such a case, the averaged darker coating shows evidence that a sufficient monomer load is gathered within a 100~mM concentrated solution.\\[3pt]
The possibility to grow CPDs under such configuration at a 100~mM monomer load has been evidenced by increasing the ion content in the drop (Fig.~\ref{fig:fig3}h--l).
At the standard level of 1~mM concentration for EmimOTf (Fig.~\ref{fig:fig3}h), the highest contrast is observed.
Gradual concentration increase up to 10~mM of the ionic liquid content induces the darkening of the overall coating (Fig.~\ref{fig:fig3}h--l), suggesting that the involvement of OTf\textsuperscript{--} as a doping counter-anion rather than an electrolyte is the limiting factor for densifying the electropolymerized PEDOT:OTf coating.
From 7.5~mM concentration (Fig.~\ref{fig:fig3}k) and more noticeably at 10~mM (Fig.~\ref{fig:fig3}l), formations of CPDs are observed with their nucleation point on the edge of the drop.
These last results validate that spots and CPDs can coexist with a single unipolar voltage growth between two parallel plates and that 100~mM of monomers are sufficient to promote CPD growths at multiple points in the cavity.\\

\subsection{Growth/Nucleation Balance with and Electrode Array Size and Density}

\begin{figure}[!h]
  \centering
  \includegraphics[width=1\columnwidth]{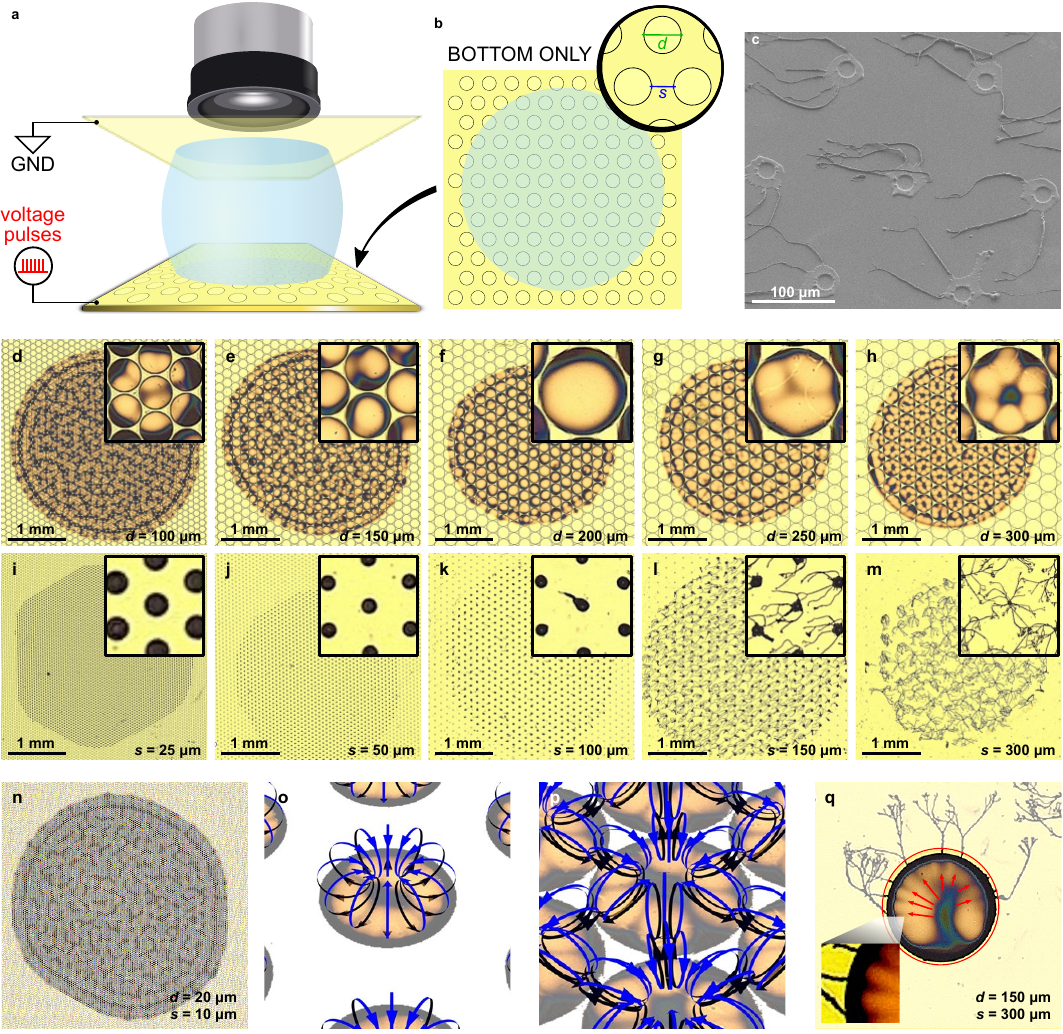}
  \caption{\textbf{The Electrodes' Geometry as the Main Factor Determining the Transition between 2D Spots and 3D Dendrites $\vert$}
  \textbf{a,} Schematic illustrating the geometry of the bottom electrode as the main variable studied in this figure. 
  Each electropolymerization was conducted using the same voltage and chemical composition (unipolar voltage, pulse~: 5~V, \textit{f}~= 1~kHz, 1~mM EmimOTf, 100~mM EDOT, 100~mM BQ, during 60~s in PC). 
  \textbf{b,} The geometry of the bottom electrode consists of an array of holes with diameter \textit{d} and spacing \textit{s}, etched in the Parylene C. 
  \textbf{c,} SEM picture of an array of holes (\textit{s}~= 150~$\upmu$m, \textit{d}~= 20~$\upmu$m) after electropolymerization and drying of the surface.
  \textbf{d--h,} Morphological evolution of the pattern with the diameter of the holes (\textit{s}~= 10~$\upmu$m). 
  \textbf{i--m,} Morphological evolution of the pattern with the spacing of the holes (\textit{d}~= 20~$\upmu$m). 
  \textbf{n,} High-contrast picture of the pattern obtained with the smallest hole diameter and spacing, revealing a macroscopic organization similar to that observed with unpatterned electrodes. 
  \textbf{o--p,} Schematics illustrating the hypothetical flow distribution for both a large spacing (\textbf{o}) and a small spacing (\textbf{p}) configurations. 
  \textbf{q,} Optical microscope zoom-in on a single hole featuring streaks, each corresponding to a CPD, highlighting the link between electroconvection and dendritic growth.
  }
  \label{fig:fig4}
  \end{figure}

To study the impact of an electrode array geometry, the following experiments were performed with the same unipolar voltage pulses and low ion concentration.
A Parylene C passivation layer is coated on the anode at the bottom, to evaluate the impact of various honeycomb electrode arrays on the morphogenesis of PEDOT patterns or their CPDs (Fig.~\ref{fig:fig4}a,b).
Such arrays have been systematically studied with circular openings of different diameters \textit{d} and spacings \textit{s} (Fig.~\ref{fig:fig4}b), to quantify the limitation of the physical boundary of the electrochemical cavity, to observe a morphological transition between a growth regime resembling the free-standing wire setup and one resembling the parallel plate electrode.\\[3pt]
Forests of CPDs have successfully been obtained (Fig.~\ref{fig:fig4}c), but not systematically and only with specific \textit{s} and \textit{d} conditions.
Microscope pictures for all electrode geometries are provided as supplementary material (Fig.\ref{fig:figS1}).
Both geometrical parameters have distinctive effects on the electrogenerated growths.\\[3pt]
Specifically, arrays featuring narrow spacings expose their gold surface in a major fraction of their total area: as a consequence, the electrogenerated growths resemble the spot patterns which are generated with plain gold plates (Fig.~\ref{fig:fig4}d--h).
However, the spot organization is controlled by the periodicity of the array in the case the opening diameter \textit{d} is larger than the size of the electrochemical cavity (\textit{e}~= 150~$\upmu$m).
When \textit{d}~= 100~$\upmu$m (Fig.~\ref{fig:fig4}d), a single opening cannot contain a whole convective cell, so the pattern organization displays a similar randomness as in the case where the Parylene C passivation layer is not coating the anode.

When \textit{d}~= 150~$\upmu$m (Fig.~\ref{fig:fig4}e), an apparent randomness remains in the spot pattern organization, but the center of each opening is systematically lighter than on its edge.
When \textit{d}~= 200~$\upmu$m (Fig.~\ref{fig:fig4}f), the spot organization is controlled both globally and locally: except on the drop edge, all openings are light at their center and dark on the Parylene C edge.
When \textit{d}~= 250~$\upmu$m (Fig.~\ref{fig:fig4}g), the openings become substantially larger than the size of the cavity \textit{e}, so multiple cells can coexist within a single opening, creating a slightly darker area at the center of the opening.
When \textit{d}~= 300~$\upmu$m (Fig.~\ref{fig:fig4}h), the openings are twice the size of the cavity \textit{e}, generating a suborganization of six cells within each opening, where the crystal orientation of the electrogenerated spots matches the lattice of the Parylene C openings, suggesting interactions between convective cells on neighboring openings.
These results suggest that by the size of the electrodes in an array, with respect to the size of the electrochemical cavity, growth randomness can be controlled at different scales.\\[3pt]
When the electrode area is the smallest, a smaller fraction of the total area is opened to the environment: as a consequence and in case the number of openings is sufficiently sparse, CPDs grow as obtained with a free-standing wire setup (Fig.~\ref{fig:fig4}i--m).
When \textit{s}~= 50~$\upmu$m or lower (Fig.~\ref{fig:fig4}i--j), the openings are dark with a coating exclusively confined within the cavity.
When \textit{s}~= 100~$\upmu$m (Fig.~\ref{fig:fig4}k), local defects start to appear in some of the openings, with dark material overflowing the Parylene C passivation layer.
When \textit{s}~= 150~$\upmu$m (Fig.~\ref{fig:fig4}l), all openings feature multiple filamentary growths.
When \textit{s}~= 300~$\upmu$m (Fig.~\ref{fig:fig4}m), the filaments seem more dendritic, with multiple branches on each dendrite and multiple dendrites on each opening.
These results suggest that by the pitch of the electrodes in an array, with respect to the diameter of the electrodes, CPDs can start to grow vertically.\\[3pt]
Other observations have been made on this set of experiments.
In the case of small and narrow openings, (Fig.~\ref{fig:fig4}n and Fig.~\ref{fig:figS1}a1 for the raw image), the total opening area is too large to promote formation of CPDs and openings are not wide enough to promote high contrast: all spots seem to be coated uniformly.
However, enhanced contrast of the original image as displayed in Fig. \ref{fig:fig4}n shows that even under such severe geometrical boundaries, the convective randomness is still persistent and imprints on the anode surface.
The link between dendrites and electroconvective vortices is still unclear, but it is hypothesized that the flow inside the openings is a key factor. 
The hypothetical flow occurring inside openings with large spacings is depicted in Fig.~\ref{fig:fig4}o based on experimental observations with tracer particles.\cite{Baron2026a} 
Vortices may form a ring ascending near the edges of the openings and descending at the center. 
This would leave a more stable area at the center where the PEDOT may primarily deposit. 
The pattern, as well as the flow structure, would be uniform in all radial directions. 
Meanwhile, when the openings are close to each other (Fig.~\ref{fig:fig4}p), it is hypothesized that the flow inside an opening communicates with its neighbor along the armchair directions of the honeycomb structure. 
This coupling may result from an overlap of the diffusion layers of both openings. 
Since the flow is concentrated along these radial axes, PEDOT deposition may be maximal along the angles between the armchair directions.
An experimental evidence of the importance of these vortices in the growth of CPDs is presented in Fig. \ref{fig:fig4}q.
The boundaries of different convective cells condition both the darkening of the anode between two spots and the nucleation of a single CPD on the edge of the gold/Parylene C interface, as pointed out by the red arrows in Fig. \ref{fig:fig4}q.
Therefore, one can affirm that electroconvection is controlling both the growth of CPDs (as shown in water by Baron \textit{et al.} by the impact of the viscosity on the charged particle motion)\cite{Baron2026} but also their nucleation.\\
  
\subsection{Effect of Electrode Array Alignment on the Nucleation and Growth Morphology}
\begin{figure}
  \centering
  \includegraphics[width=1\columnwidth]{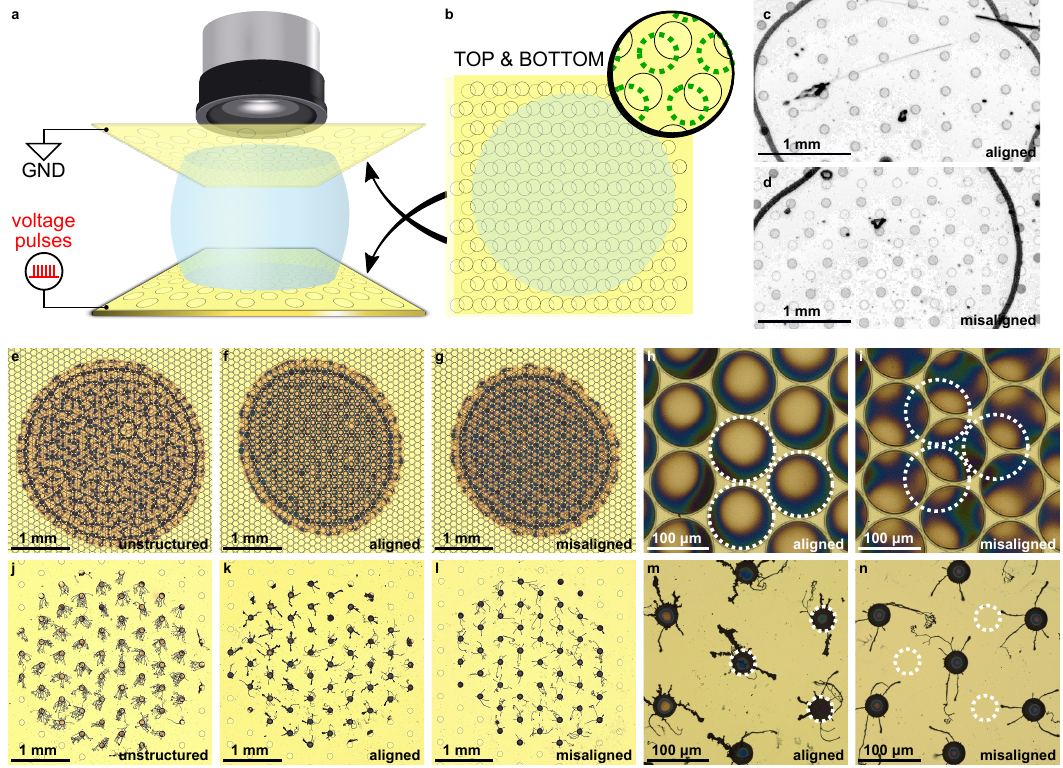}
  \caption{\textbf{Effect of the Patterning of Both Counter and Working Electrodes on the Growth of PEDOT Structures. $\vert$}
  \textbf{a,} Schematic illustrating the geometries of the bottom and top electrodes as a first variable studied in this figure. 
  \textbf{b,} Illustration of the second variable of interest: the alignment of the openings etched in the Parylene C covering both electrodes. 
  \textbf{c--d,} Optical microscope pictures of the system after the droplet is placed on the bottom electrode and covered on top by the top electrode, in two configurations (\textbf{c} shows in-phase openings, \textbf{d} shows out-of-phase openings). 
  \textbf{e--n,} Microscope pictures of the deposition obtained in different experimental conditions, using the same set of voltage parameters and chemical composition (unipolar voltage, pulse~=~5~V, \textit{f}~= 1~kHz, 1~mM EMIMOTf, 100~mM EDOT, 100~mM BQ, during 60~s in PC). 
  Specifically: \textbf{e,} Macroscopic view of a reference system with small spacing (\textit{s}~= 10~$\upmu$m, \textit{d}~= 100~$\upmu$m) and an unpatterned top electrode. 
  \textbf{f--g,} Macroscopic views of the deposition for the same electrode geometry, when openings are in-phase (\textbf{f}) and out-of-phase (\textbf{g}). 
  \textbf{h--i,} Close-up views of the pattern formed inside the in-phase openings (\textbf{h}) and inside the out-of-phase openings (\textbf{i}). 
  \textbf{j,} Macroscopic view of a reference system with a large spacing (\textit{s}~= 300~$\upmu$m, \textit{d}~= 100~$\upmu$m) and an unpatterned top electrode. 
  \textbf{k--l,} Macroscopic views of the deposition for the same geometry, when openings are in-phase (\textbf{f}) and out-of-phase (\textbf{g}). 
  \textbf{m--n,} Close-up views of the dendrites grown from the in-phase openings (\textbf{m}) and from the out-of-phase openings (\textbf{n}).
  }
  \label{fig:fig5}
  \end{figure}

To electrochemically interconnect an array of electrodes with another array of electrodes, the system complexity increases naturally by the higher number of geometric parameters.
To study the impact of the electrode alignment on the growth, top and bottom electrodes have been structured with the same opening diameter \textit{d} and spacing \textit{s} to keep the same periodicity between the bottom anode and the top cathode (Fig.~\ref{fig:fig5}a--b). 
Different systems have been studied under two different configurations: "aligned" as the openings on the cathode vertically align with the ones on the anode (Fig.~\ref{fig:fig5}c), and "misaligned" as the openings on the cathode are quasi-equidistant to three openings on the anode (Fig.~\ref{fig:fig5}d).\\[3pt]
When the cathode is either aligned (Fig.~\ref{fig:fig5}f) or misaligned (Fig.~\ref{fig:fig5}g) with the anode, spots are more ordered than without a structured cathode (Fig.~\ref{fig:fig5}e).
A closer observation of these electrogenerated growths indicates that the position of the cathode controls the placement of the spots on the anode (Fig.~\ref{fig:fig5}h--i).
This relates to previous results obtained in water, showing that the structuration of the cathode allows conditioning the patterning of ordered structures on the anode when the cathode is lithographically structured as a hexagonal mesh of gold lines.\cite{Baron2026b}
In the present case, both the anode and the cathode are geometrically identical.
Therefore, it indicates that both electrodes are not equivalent and that the electrodes' geometries also control the positions of electroconvective cells with unipolar voltage stimulation.\\[3pt]
In case of the formation of CPDs, both the structuration of the top array and its alignment with the bottom one have an impact on the growth.
In both cases (Fig.~\ref{fig:fig5}k--l), structuring the cathode leads to fewer CPD growths than in the previous case where the cathode is not structured with a Parylene C coating (Fig.~\ref{fig:fig5}j).
In the case where the openings are aligned, two different morphologies of dendrites coexist within the same opening (Fig.~\ref{fig:fig5}k,m): thick-and-short ones, and thin-and-long ones.
In case of misaligned openings, only thin-and-long dendrites are visible on the bottom substrate (Fig.~\ref{fig:fig5}l,n), sometimes terminated by several smaller ramifications.
One possible reason for such differences is that in the case of the misaligned configuration, growths are directional exclusively between triplet of close neighbors, while in the case of the aligned configuration, the orthogonal growth competes with the six next neighbors.
These experiments highlight the impact of the counter-electrode's geometry on the flow, which in turn affects how the dendrites develop, and the important role of the alignment. 
These elements are important geometric variables to take into account for the integration of vertically aligned CPDs on an array of electrodes.   
 
\newpage

\subsection{Dynamic Signal Evolution from an Array of Vertically-Aligned Conducting Polymer Dendrites}
\begin{figure}[!h]
  \centering
  \includegraphics[width=0.97\columnwidth]{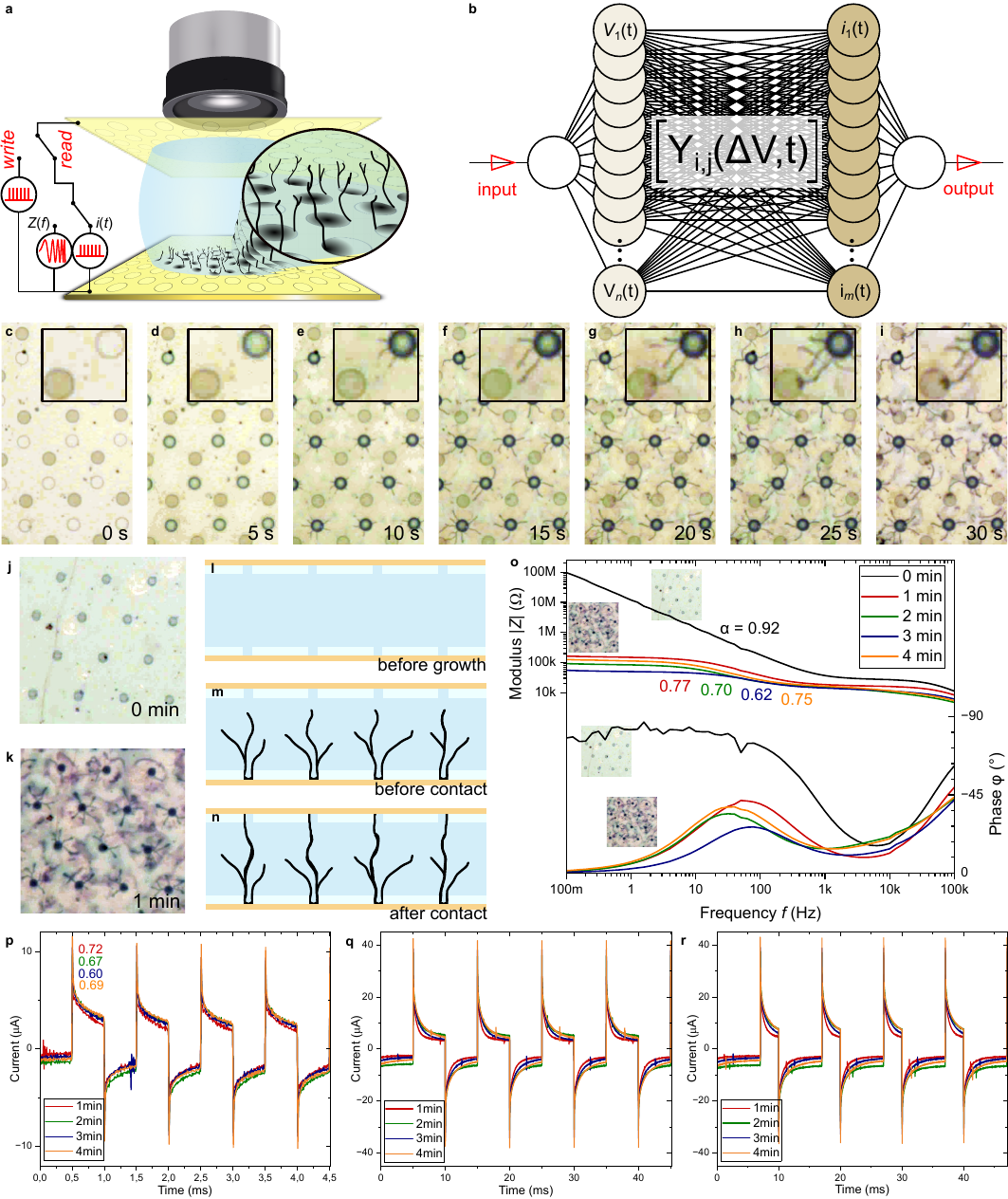}
  \caption{\textbf{Electrical Characterization of an Array of Vertical Conducting Polymer Dendrites $\vert$} 
  \textbf{a,} Schematic of the electrically programmable two-array system, growing dendrites on the bottom array. 
  \textbf{b,} System as n input nodes connected to m output nodes with their corresponding admittance Y\textsubscript{i,j} dependent on both elapsed time and past voltages. 
  \textbf{c--i,} Optical microscope pictures of a misaligned configuration (\textit{s}~= 300~$\upmu$m, \textit{d}~= 100~$\upmu$m) taken at different times as dendrites are grown from the bottom holes (5~V unipolar pulses, 1~kHz, 1~mM EmimOTf, 100~mM EDOT, 100~mM BQ in PC). 
  \textbf{j--k,} Top view on which impedance is measured (\textit{s}~= 260~$\upmu$m, \textit{d}~= 50~$\upmu$m, aligned holes) before (\textbf{j}) and after (\textbf{k}) electropolymerization (5~V unipolar pulses, 1~kHz, 1~mM EmimOTf, 100~mM EDOT, 100~mM BQ in PC).
  \textbf{l--n,} Schematics illustrating the three stages of CPD growth between two planar electrodes.
  \textbf{o,} Impedance measurements before and after dendrites' growth.
  \textbf{p--r,} Current responses to square wave voltages (\textbf{p}: 50~mV, 1~kHz, dc~= 50\%. \textbf{q}: 200~mV, 100~Hz, dc~= 50\%. \textbf{r}: 200~mV, 100~Hz, dc~= 30\%).
  }
  \label{fig:fig6}
\end{figure}

\begin{figure}[!h]
  \centering
  \includegraphics[width=1\columnwidth]{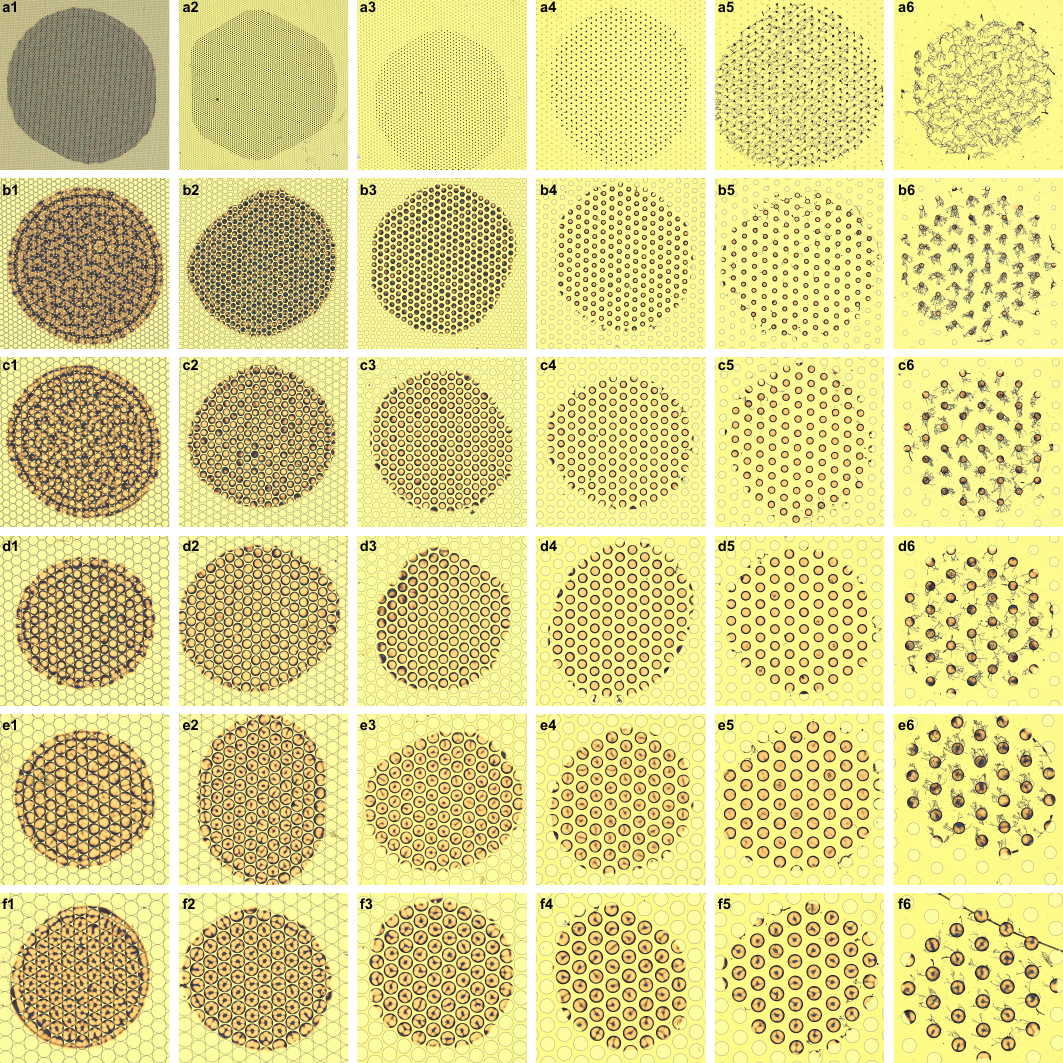}
  \caption{\textbf{Microscope Pictures of the 6$\times$6 Patterns Generated from Different Array Geometries $\vert$ }
  All depositions were realized with the same voltage, solution (unipolar voltage, \textit{V}\textsubscript{p}~= 5~V, \textit{f}~= 1~kHz, 1~mM EmimOTf, 100~mM EDOT, 100~mM BQ, during 60~s in PC) and gold substrate.
  Diameter \textit{d}~= 20~$\upmu$m (\textbf{a.}), 100~$\upmu$m (\textbf{b.}), 150~$\upmu$m (\textbf{c.}), 200~$\upmu$m (\textbf{d.}), 250~$\upmu$m (\textbf{e.}) and 300~$\upmu$m (\textbf{f.}).
  Spacing \textit{s}~= 10~$\upmu$m (\textbf{.1}), 25~$\upmu$m (\textbf{.2}), 50~$\upmu$m (\textbf{.3}), 100~$\upmu$m (\textbf{.4}), 150~$\upmu$m (\textbf{.5}) and 300~$\upmu$m (\textbf{.6}).
  } 
  \label{fig:figS1}
\end{figure}

In the previous experiments, the impact of the voltage, the concentrations and the electrode geometry on the morphologies of electrogenerated PEDOT growth was studied on a 2D substrate.
These morphologies were characterized by optical microscopy \textit{post mortem}, after rinsing and drying them: although these steps do not alter the morphology of spot patterns, the collapsing of vertically aligned CPDs does not help to capture information other than their dendricity and branch thickness (Fig.~\ref{fig:fig6}a).
To assess their growth orientation and study their electrical connectivity \textit{in operando} (Fig.~\ref{fig:fig6}b), the following data were recorded with CPDs continuously immersed in their wet environment.\\[3pt]
To the best of our knowledge, all CPD growths reported in the literature were performed with bipolar voltage waveforms.
Under unipolar voltage, only one electrode can oxidize the EDOT monomer.
To confirm that CPDs can reach the cathode when growths are driven by unipolar voltage pulses, multiple time-lapses of a growth video with the misaligned configuration are displayed in Fig. \ref{fig:fig6}c--i (the image contrast is enhanced for better clarity).
In the image \ref{fig:fig6}c, the openings of the anode lower substrate appear clearer than the ones of the cathode upper substrate.
After a few seconds of growth (Fig.~\ref{fig:fig6}d), the edge of the openings starts to darken, and dendrites start to grow on the periphery after ten seconds (Fig.~\ref{fig:fig6}e).
They expand (Fig.~\ref{fig:fig6}f) and seem to reach the cathode in about 20 seconds (Fig.~\ref{fig:fig6}f).
The physical connection is further confirmed by the local thickening of the dendrites apex on the apparent contacts with the cathode openings (Fig.~\ref{fig:fig6}g,h). 
The image further confirms that an anode opening can connect simultaneously with multiple of its three closest cathode opening neighbors, and that a cathode opening can receive simultaneously different dendrites from their three closest anode opening neighbors.
Although all openings are always electrochemically interconnected (the schematic in Fig. \ref{fig:fig6}b holds true at all times for dynamical signals), the possibility of physically connecting openings beyond their nearest neighbors appears to be limited in this demonstration.
However, it is expected that colloidal growth media with intrinsically more disorder than a single phase liquid solution could enhance the physical interconnection of openings beyond their closest neighbor shell.\\[3pt]
Attempting to measure the electrochemical interconnectivity during the first seconds before contact without perturbing the growth proved unsuccessful.
Unlike previous experiments performed with more massive CPDs on a free standing wire setup under bipolar voltage pulses,\cite{Baron2024} the interruption of the growth waveform to intermittently read the impedance state at their juvenile state perturbs the growth.
Instead, multiple impedance and current characterizations were performed at a more mature state after several minutes when multiple CPDs contact both planes (Fig.~\ref{fig:fig6}j--r). 
In this following experiment, performed in the aligned configuration, most of the growth is performed within the first minute (Fig.~\ref{fig:fig6}j,k): before growth, the arrays are free of any PEDOT polymer (Fig.~\ref{fig:fig6}l), the juvenile states of CPDs before contact (Fig.~\ref{fig:fig6}m) are assumed to occur globally before the end of the first minute, to reach a more mature state after contact (Fig.~\ref{fig:fig6}n) with the first closest neighbors.
To evaluate if the system can evolve further after connecting the first closest neighbor within the first minute, electrochemical impedance spectroscopy and transient current characterization were performed after interrupting the growths every minute.\\[3pt]
As observed with the free-standing wire setup in water, most of the impedance change occurs by the coating of the first PEDOT oligomers on the gold interface (Fig.~\ref{fig:fig6}o).\cite{Baron2024}
The impedance of the system before growth and after 1~min differ by orders of magnitude (Fig.~\ref{fig:fig6}o), confirming that the presence of these small PEDOT fibers has a significant impact on the signal to be transported through the electrochemical cavity.
In all measurements recorded after the first minute, a low frequency impedance plateau is observed, which is characteristic of an electrical short between the anode and the cathode: no faradaic process is assumed, as the same solution stressed under the same conditions did not display such a plateau before growth (Fig.~\ref{fig:fig6}o).
Over the minutes, a variation of the depth of this plateau is observed: 137$\pm$31~k$\Omega$ at 1~min, 77$\pm$17~k$\Omega$ at 2~min, 44$\pm$13~k$\Omega$ at 3~min and 106$\pm$23~k$\Omega$ at 4~min (Fig.~\ref{fig:fig6}o). 
The fact that the system's impedance continuously decreases between one and three minutes shows that the first contact does not disable further evolutions of the system: despite the electric short caused by the first dendrites, the low resistance caused by such small fibers does not prevent the anode to oxidize EDOT further.
On the video, the formation of new dendrites after the first minute is not evidenced: the cause for such decrease of the impedance could potentially be the thickening of the already existing CPDs.
We also noticed that the system's impedance returns to a higher state at 4~min.
Possibly due to mechanical rupture of some CPDs, this sign of fatigue can be used in a relevant way to reset the interconnection between inputs and outputs in an evolving system.\\[3pt]
Dynamical current traces were also recorded at different pulse regimes for each mature state captured by impedance spectroscopy, displayed earlier in Fig. \ref{fig:fig6}o (Fig.~\ref{fig:fig6}p--r).
With either low voltage amplitude (Fig.~\ref{fig:fig6}p), high voltage amplitude (Fig.~\ref{fig:fig6}q) or asymmetric charge (Fig.~\ref{fig:fig6}r), no significant differences were identified in the temporality of the current decays.

Particularly obvious for a low voltage amplitude, such current decays are not ruled by ideal Debye relaxations: both charge and discharge decay non-exponentially (Fig.~\ref{fig:fig6}p).\cite{baron2026bioelectrical}
Yet, the nature of such behavior does not seem to be of a complexity as high as the one obtained from massive single CPDs.\cite{Baron2024,Baron2025}
If multiple relaxations are to be attributed to multiple CPDs growing from the same anode, their individual relaxations contribute collectively to one single global dynamics.
Fitting of both impedance spectra and current traces to a 1/\textit{G}+\textit{R}~|~CPE(\textit{Q},\textit{$\alpha$}) circuit model \cite{Baron2024,Baron2024a} reveals a clear evolution of the constant-phase-element (CPE) attenuation coefficient $\alpha$ with the CPD growth (Table \ref{table:tab1}):

\begin{table}[!htbp]
\centering
\caption{Equivalent Circuit Fitting Parameters (by \textit{Z}(\textit{f}) | by \textit{i}(\textit{t}))}
\begin{tabular}{*5c}
  \toprule
  Growth Time (min) & 1/\textit{G} (k$\Omega$) & \textit{R} (k$\Omega$) & \textit{Q} (nF$\cdot$s\textsuperscript{$\alpha$-1}) & $\alpha$ \\ 
  \midrule
  0 & 27.6 | ------ & 1.43 $\times$10\textsuperscript{6} | ------ & 15.7 | ------ & 0.92 | ------ \\
  1 & 16.0 | 16.9 & 116 | 181 & 209 | 13.9 & 0.77 | 0.72 \\
  2 & 13.2 | 13.1 & 68.5 | 86.1 & 634 | 339 & 0.70 | 0.67 \\
  3 & 13.4 | 12.6 & 34.7 | 64.4 & 551 | 253 & 0.62 | 0.60 \\
  4 & 14.1 | 12.2 & 95.6 | 129 & 446 | 106 & 0.75 | 0.69 \\
  \label{table:tab1}
\end{tabular}
\end{table}

Interestingly, one can observe that both fittings in the time and frequency domains show that from 2~min, an increased time for the growth leads to the diminishing of the pseudo-capacitance \textit{Q}.
This was never characterized before in the case of a single massive and free-standing CPD, as their size led to a too conductive electronic transport relative to the ionic transport after completion (the residual ionic decay was too weak for accurate fitting).\cite{Baron2024,Baron2025}
Here, by confining the CPDs between two plates, their characteristic dimensions allow to balance the electronic and ionic impedances upon contact. This enables the monitoring of their incremental evolution via transient signals, even after the first dendritic filaments bridge the gap.
Except in the last case at 4~min, the decrease of the attenuation coefficient $\alpha$ is also noticed.
The concomitant decreases of \textit{R}, \textit{Q} and $\alpha$ indicate that a gradual evolution of the signal transport from capacitive to resistive is still to be expected even after contact, so that calibrating the array is not intrinsically limited by the growth kinetics of the polymer through the cleft.\\[3pt]
These electrical characterizations of integrated CPDs on a single electrolyte layer deliver convincing promises for their further integration in a multilayered system.
With appropriate compositions, sizes and alignments, multiple dendrites can grow between two arrays.
Their morphogenesis can program the transport of dynamical signals at their juvenile states (non contacting) but their dynamics are also readable at a more mature state, after contact, to carry information.
This demonstrates that the programmability of these objects is not limited by the electrical conductivity of the material when spike signals are used to write and read an array of CPDs.
Akai-Kasaya and coworkers showed that filamentary switching could be a relevant programming framework to exploit such evolving objects as memristors in a wetware.\cite{AkaiKasaya2020,Hagiwara2023}
Thanks to integrating them as an array of vertically-aligned CPDs, these new results show that reading them with spike voltages in AC or accessing their impedance rather than reading their conductance in DC is also possible and may even encode more information on a higher dimensional space, exploiting their rich morphology as a resource for information storage and processing.
  
\section{Conclusions}
This study demonstrates the possibility of integrating CPD morphogenesis in wet-film devices.
To promote the mass transfers of conducting polymer particles required for the growth of dendrites, propylene carbonate is used as a low toxicity and low volatility organic solvent.
Similar evolutions as those performed in water are observed with an applied voltage and from the same electroactive precursors.
The larger electrochemical stability of the solvent and higher affinity with the monomer precursors allowed us to extend the voltage and concentration boundaries of the process to further understand the underlying growth mechanism.
Moreover, their inherent disorder is not entirely caused by stochastic particle motion in the fluid during growth but relies on a disordered nucleation, induced by electro-convection with a DC electric field: so frequency-dependent dendrite fractality and nucleation randomness are governed by two separate statistical mechanisms.
The possibility of increasing the monomer density in a wet film allowed a sufficiently large content of precursors to observe multiple dendritic growths on an array.
Electrode size, pitch and alignment are confirmed as crucial parameters to trigger the electrochemical growth of a conducting polymer from 2D to 3D in a wet cavity.
Despite the convenience of using a liquid as an interlayer medium, enabling evolving electronics to exploit CPD morphogenesis should rely on a technology where the stacking of multiple electrode arrays is highly controlled to not bias programmability.
Growths that reached maturity were evidenced to completely cross the cavity, without preventing further evolutions of the electrochemical impedance of the system, leading to multiple dendrites interconnecting two arrays on different contacts and with different directions.
Given the benefits of CPDs for both sensing complex environments,\cite{Lengefeld2026} and classifying their activity,\cite{Cucchi2021} this study shows the feasibility for them to grow from a 2D substrate with a high throughput to achieve higher complexity systems.
As an earlier work showed that parallelizing CPD growths could be a way to program conductometric sensors to enable evolutive functions in an electronic nose,\cite{Baron2026c} a next challenge will be to achieve a monolithic integration of sensing arrays with such wetware.\\[3pt]
Beyond their role as dynamic interconnects, the demonstrated CPD arrays offer a potential platform for reservoir computing. Their computational relevance comes from the combination of morphology-dependent connectivity, evolving impedance and non-exponential electrical transients, which could enable input signals to be transformed into history-dependent electrical outputs. In this framework, morphogenesis could serve not only to fabricate a computational substrate but also to configure its dynamical response, acting as a means of storing physical information through the formation of complex structures \cite{pismen2020}. Since the evolving conductive network is formed electrochemically from a relatively simple polymer precursor, it could potentially reduce reliance on fabrication infrastructure and rare materials \cite{Alluhaibi2025}. Therefore, having demonstrated the feasibility of interconnecting two electrode arrays through 3D dendritic growth, this study opens the way to exploring the computational potential of such 3D networks, particularly in frameworks like reservoir computing.

\section*{Acknowledgments}
The authors thank the French National Nanofabrication Network \href{https://www.renatech.org/en/}{RENATECH} for financial support of the IEMN cleanroom. 
We thank the IEMN cleanroom staff for their advice and support. 
This work is funded under the ANR-JCJC "Sensation" project (grant number: \href{https://anr.fr/Projet-ANR-22-CE24-0001}{ANR-22-CE24-0001}).
The authors acknowledge the support of the Erasmus+ programme, administered through AGH University of Krakow, which enabled the research traineeship at IEMN.
We are deeply grateful to the French 'GdR' Network \href{https://defie-cnrs.fr/}{DEFIE} for the very inspiring exchanges on "Green Electronics".

\section*{Competing Interests}
The authors declare no competing interests.

\bibliographystyle{natsty-doilk-on-jour}  
\bibliography{ref}  

\newpage

\section*{Supplementary Materials: Sprouting Fields of Conducting Polymer Dendrites as Evolving Organic Electrochemical Frameworks to Interconnect Electrode Arrays}

\beginsupplement

\end{document}